\documentclass{article}
\usepackage[T1]{fontenc} % add special characters (e.g., umlaute)
\usepackage[utf8]{inputenc} % set utf-8 as default input encoding
\usepackage{lipsum}

\newcommand\blfootnote[1]{%
  \begingroup
  \renewcommand\thefootnote{}\footnote{#1}%
  \addtocounter{footnote}{-1}%
  \endgroup
}

\def\authorname{E. Manilow and B. Pardo}

\def\papertitle{Bespoke Neural Networks for Score-Informed Source Separation}

\usepackage{ismir-lbd}

\usepackage{lineno}
\title{\papertitle}

\twoauthors
 {Ethan Manilow} {Interactive Audio Lab \\ Northwestern University, Evanston, IL}
 {Bryan Pardo} {Interactive Audio Lab \\ Northwestern University, Evanston, IL}

\begin{document}

\maketitle
\begin{abstract}
In this paper, we introduce a simple method that can separate arbitrary musical instruments from an audio mixture. Given an unaligned MIDI transcription for a target instrument from an input mixture, we synthesize new mixtures from the midi transcription that sound similar to the mixture to be separated. This lets us create a labeled training set to train a network on the specific bespoke task.
When this model applied to the original mixture, we demonstrate that this method can:  1) successfully separate out the desired instrument with access to only unaligned MIDI, 2) separate arbitrary instruments, and 3) get results in a fraction of the time of existing methods. We encourage readers to listen to the demos posted here: \url{https://git.io/JUu5q}.
\end{abstract}
\section{Introduction}\label{sec:introduction}
Source separation in the musical context is the task of extracting the audio of one or more target instruments from an audio mixture containing multiple concurrent instruments. Today's top musical source separation methods all rely on deep learning models trained on large sets of labeled training data. All machine learning methods rely on the assumption that the distribution of the training data matches the distribution of data out in the real world. When this assumption is broken, the performance of a model suffers. It is not possible to guarantee the system was trained on data relevant to the current separation task using this paradigm.

In this paper, we introduce a method that explicitly avoids the problem of mismatch between the training and real world data, but creating a bespoke training set specifically for the task at hand. Inspired by Deep Image Prior~\cite{ulyanov2018deep},
% where a network is trained to reconstruct exactly one image from an input of noise,
our method aims to separate out exactly one source from exactly one mixture by training a separation model on a dataset created specifically to be similar to the mixture that requires separation. The resulting bespoke network is trained specifically to solve the task at hand.

\section{Bespoke Neural Networks}\label{sec:bespoke}

This paper introduces \textit{Bespoke Neural Networks} for musical source separation, where a network is trained to only separate exactly one target source from exactly one mixture. We assume the user can provide an unaligned MIDI transcription for the target source in the mixture. We create synthesized audio from the MIDI, and synthesize the source in one or more ways~\cite{manilow2019cutting, miron2017generating}. We then apply many types of augmentations to the synthesized source, and mix it together with background audio. We describe the specifics of augmentation, the background audio, and mixing in the next section. Using the known synthesized source as the ground truth source label, we train a small neural network to ``overfit'' to these local augmented examples as surrogate to the actual mixture. After small number of training iterations, we then use the trained network on the original mixture to retrieve an estimate of the desired source.

Because we assume that some transcription of the desired source is available \textit{a priori}, we say that this method is score-informed. We note that unlike traditional methods for score-informed separation~\cite{ewert2014score}, our method does not need perfectly aligned score data. Other deep net methods have been proposed for score-informed separation~\cite{gover2019score} or for separating and transcribing sources~\cite{manilow2020simultaneous}, however they are limited in types of source types they can separate and require large amounts of time and data to train. 

There are many advantages to this method over existing methods. 
First, our method only requires an unaligned MIDI transcription of a target source, which we assert is a much less stringent requirement than the large corpus of isolated source data needed by other methods.
% This side-steps the issue of having to gather a large corpus of isolated source data, which may be difficult or impossible if one wants to separate a source .
% which we assert is easier to acquire than more isolated source data to train a model for a new source type.
% First, our method is able to train in instances where ground truth audio source data is unavailable. If given a coarse transcription, our method is able to separate by synthesizing its own labels that extrapolate to the original mix. We assert that obtaining one unaligned transcription is a much more lightweight process than obtaining audio source data to separate a particular source.

Second, our method is not expected to model a large distribution of data in the hopes that it will generalize well to every unseen musical mixture. It is only expected to generalize to exactly one source within exactly one mixture. This gives our method the flexibility of separating a myriad of diverse source types--any source type that can be synthesized. Compare this to many existing methods, which are limited by whatever coarse set of source types might be available in a large enough numbers to train a model with (\textit{e.g.,} ``Vocals'', ``Bass'', ``Drums'', \& ``Other''). 

Finally, because the network is designed to ``overfit'', we drastically reduce the its size and number of training steps, making the proposed method much faster. Our method is able to produce suitable results in \textit{minutes}, instead of hours or days for typical source separation networks. This makes one-off bespoke training practical to use.

% In the following section we discuss two demo outputs from the proposed method. We encourage readers to listen to the demo files: \url{https://git.io/JURcA}

\section{Demos}

In this section we describe the process of creating two demo excerpts using bespoke networks. Both demo clips were created by training mask inference network with a 2 BLSTM layers with 300 units each and a dropout at 30\% zeroing probability applied to the second layer. After the BLSTM layers, a fully connected layer with a sigmoid activation function estimated a mask. The whole network was trained to minimize the truncated Phase Sensitive Approximation (tPSA)~\cite{erdogan2015phase} objective function.
% Every iteration, augmentations to the mixture and sources were applied randomly as described below.

Each training step, mixes were created randomly such that a gain for the background and synthesized target were chosen from a uniform distribution in the interval $[-12, 6]$ dB. The background and target were then mixed together. The details of the background audio differ for each demo, below. Dynamic range compression was applied to the mix with a reduction ratio chosen randomly from the set $\{2, 4, 8, 12, 16, 20\}$ and finally the training mixes were peak normalized. All models were trained for 2,000 steps and training times are reported using a Nvidia V100 GPU.
% We were unable to find ground truth isolated source data for the excerpts below, and thus cannot provide evaluations using standard metrics.

For these demos, we specifically chose to isolate sources that would be difficult or impossible to separate using existing deep learning source separation systems. Listen to the mixes, estimated sources, the synthesized source training data, and output from competing methods at our demo website: \url{https://git.io/JUu5q}

\subsection{Take On Me -- A-ha}

The first example that we demonstrate is an excerpt from the 1984 hit ``Take On Me'' by A-ha. We chose to isolate the famous synthesizer melody at the beginning of the song. The rest of the mixture contains a synthesizer countermelody, a synthesizer ``pad'', a bass, and a drum track. 
% We note that the synthesizer melody, countermelody, and pad are all considered the ``other'' source in Spleeter and other source separation

This example was created by synthesizing just one example of the target synth melody with a synthesizer that sounded loosely similar to the target source. We then created training mixes by summing the synthesized target together with the input mixture. Training took less than 5 minutes.

% A gain for both the input mixture and synthesized target chosen from a uniform distribution in the interval $[-12, 6]$ dB. Dynamic range compression was applied with a reduction ratio chosen randomly from the set $\{2, 4, 8, 12, 16, 20\}$ and finally the training mixes were peak normalized. Training ran for 2000 iterations and took less than less than 5 minutes.

\subsection{Reelin' in the Years -- Steely Dan}

The second example is an excerpt from an instrumental section of the 1972 song ``Reelin' in the Years'' by Steely Dan. During this excerpt, a piano, bass, drum set, a rhythm guitar, and two lead guitars two guitars are playing simultaneously. The two guitars are highly correlated, playing a harmonized melody line in diatonic thirds. We choose to isolate just \textit{one} of the two guitars that was playing the lower part.

For this example, we found a MIDI transcription of the song online and synthesized all of the parts with 2 instrument patches for each instrument. We noticed that the transcription was incorrect in a few small but musical ways (i.e., one would not guess the MIDI was incorrect without studying the audio excerpt). To create the mixtures, we selected one patch for each instrument, and created mixes with the procedure above. Training took about 10 minutes.
% all together with a random gain selected from a normal distribution with a mean of -12 dB and std. dev. of 6 dB.
% The network was trained to isolate the target second guitar melody for 10,000 iterations which took 1hr.

\section{Conclusions and Future Work}

In this paper, we introduced Bespoke neural networks and showed how they can be used for score-informed source separation. We demonstrated our method by showcasing two songs where we were able to quickly and effectively isolate sources that would have been neglected by other source separation systems.
\blfootnote{This work has made use of the Mystic (Programmable Systems Research Testbed to Explore a Stack-WIde Adaptive System fabriC) NSF-funded infrastructure at Illinois Institute of Technology, NSF award CRI-1730689.}

In preparing these demos, we discovered that the source separation quality for a particular song is highly dependent on the augmentation strategies. We hope that future work on bespoke networks will address this problem to find a ``one-size-fits-all'' augmentation strategy and provide a scientific evaluation of the proposed method.

% \section{Acknowledgements}

% For bibtex users:
\bibliography{ISMIRtemplate}

% For non bibtex users:
%\begin{thebibliography}{citations}
% \bibitem{Author:17}
% E.~Author and B.~Authour, ``The title of the conference paper,'' in {\em Proc.
% of the Int. Society for Music Information Retrieval Conf.}, (Suzhou, China),
% pp.~111--117, 2017.
%
% \bibitem{Someone:10}
% A.~Someone, B.~Someone, and C.~Someone, ``The title of the journal paper,''
%  {\em Journal of New Music Research}, vol.~A, pp.~111--222, September 2010.
%
% \bibitem{Person:20}
% O.~Person, {\em Title of the Book}.
% \newblock Montr\'{e}al, Canada: McGill-Queen's University Press, 2020.
%
% \bibitem{Person:09}
% F.~Person and S.~Person, ``Title of a chapter this book,'' in {\em A Book
% Containing Delightful Chapters} (A.~G. Editor, ed.), pp.~58--102, Tokyo,
% Japan: The Publisher, 2009.
%
%
%\end{thebibliography}

\end{document}